\documentclass[graybox]{svmult}

\smartqed
\usepackage{mathptmx}       
\usepackage{helvet}         
\usepackage{courier}        
\usepackage{type1cm}        
\usepackage{graphicx}       

\usepackage{array,colortbl}
\usepackage{amsmath,amsfonts,amssymb,bm} 
\usepackage{multirow}
\usepackage{booktabs} 
\usepackage{pdflscape}
\usepackage[utf8]{inputenc}
\usepackage{xcolor}

\newcolumntype{H}{>{\setbox0=\hbox\bgroup}c<{\egroup}@{}}

\DeclareSymbolFont{bbold}{U}{bbold}{m}{n}
\DeclareSymbolFontAlphabet{\mathbbold}{bbold}

\usepackage{microtype} 

\usepackage[colorlinks=true,linkcolor=black,citecolor=black,urlcolor=black]{hyperref}
\usepackage{bookmark}
\makeatletter 
  \providecommand*{\toclevel@author}{999}
  \providecommand*{\toclevel@title}{0}
\makeatother

\usepackage{tikz}
\usetikzlibrary{shapes,decorations,arrows,automata,patterns,plotmarks,petri}
\usepackage{pgfplots}
\pgfplotsset{compat=1.14}
\usepackage{xcolor}

\def\be  {{\bm e}}
\def\bu  {{\bm u}}
\def\bU  {{\bm U}}

\def\by  {{\bm y}}
\def\bY  {{\bm Y}}
\def\bz  {{\bm z}}
\def\bZ  {{\bm Z}}

\def\bone  {{\bm 1}}

\def\cD   {\mathcal{D}}

\def\cG   {\mathcal{G}}
\def\cH   {\mathcal{H}}
\def\cI   {\mathcal{I}}
\def\cJ   {\mathcal{J}}

\def\cL   {\mathcal{L}}

\def\cO   {\mathcal{O}}
\def\cP   {\mathcal{P}}

\def\cV   {\mathcal{V}}

\def\EE{\mathbb{E}}

\def\II{\mathbb{I}}
\def\PP{\mathbb{P}}
\def\RR{\mathbb{R}}

\def\?{\discretionary{}{}{}}  

\def\MISE{{\rm MISE}}

\def\ISB{{\rm ISB}}

\def\IV{{\rm IV}}

\def\tr{{\sf t}}
\def\d{{\rm d}}

\newcommand{\frakv}{{\mathfrak{v}}}

\newcommand{\Var}{\mathrm{Var}}

\newcommand{\mc}{{\rm mc}}

\newcommand{\rqmc}{{\rm rqmc}}
\newcommand{\cde}{{\rm cde}}
\newcommand{\lrde}{{\rm lrde}}

\definecolor{vector}{cmyk}{0.0,0.8,1.0,1.0}  
\definecolor{tan}{cmyk}{0.30,0.50,0.60,0}
\definecolor{orange}{cmyk}{0.0,0.6,1.0,0.1}
\definecolor{names}{cmyk}{1.0,0.0,1.0,0.14}
\definecolor{pink}{cmyk}{0.0,0.8,0,0}
\definecolor{paleyellow}{cmyk}{0,0,0.6,0.0}
\definecolor{darkyellow}{cmyk}{0,0.2,1.0,0.2}

\newif\ifnotes\notestrue
\def\boxnote#1#2{\ifnotes\fbox{\footnote{\ }}\ \footnotetext{ From #1: #2}\fi}

\colorlet{pierre}{red}
\colorlet{florian}{blue!70!black}
\colorlet{TODO}{blue}
\colorlet{PERHAPS}{gray}

\newcommand{\hpierre}[1]{{}}
\newcommand{\hp}[1]{{}}

\newcommand{\florian}[1]{\boxnote{Florian}{\color{florian}#1\color{black}}}
\def\hflorian#1{}

\newcommand{\new}[1]{{#1}}

\begin{document}

\title*{Density Estimation by Monte Carlo and Quasi-Monte Carlo}

\author{Pierre L'Ecuyer \and Florian Puchhammer}

\institute{
 Pierre L'Ecuyer
 \at D\'epartement d'Informatique et de Recherche Op\'erationnelle,
     Universit\'e de Montr\'eal, Canada,
 \email{lecuyer@iro.umontreal.ca}
\and
 Florian Puchhammer
 \at Universit\'e de Montr\'eal, Canada, and Basque Center for Applied Mathematics, Spain,
     \email{fpuchhammer@bcamath.org}
}

\maketitle

\def\ourabstract{%
Estimating the density of a continuous random variable $X$ has been studied extensively in statistics, in the setting where $n$ independent observations of $X$ are given a priori and one wishes to estimate the density from that. Popular methods include histograms and kernel density estimators. In this review paper, we are interested instead in the situation where the observations are generated by Monte Carlo simulation from a model. Then, one can take advantage of variance reduction methods such as stratification, conditional Monte Carlo, and randomized quasi-Monte Carlo (RQMC), and obtain a more accurate density estimator than with standard Monte Carlo for a given computing budget. We discuss several ways of doing this, proposed in recent papers, with a focus on methods that exploit RQMC. A first idea is to directly combine RQMC with a standard kernel density estimator. Another one is to adapt a simulation-based derivative estimation method such as smoothed perturbation analysis or the likelihood ratio method to obtain a continuous estimator of the \new{cumulative density function (CDF)}, whose derivative is an unbiased estimator of the density. This can then be combined with RQMC. We summarize recent theoretical results with these approaches and give numerical illustrations of how they improve the convergence of the mean square integrated error.
}

\abstract*{
\ourabstract
}

\abstract{
\ourabstract
}

\section{Introduction}
\label{sec:intro}

In September 2015, the first author (PL) had an interesting lunchtime discussion 
with Art~Owen and Fred~Hickernell at a workshop on High-Dimensional 
Numerical Problems, at the Banff International Research Center, 
in the Canadian Rocky Mountains.  It went as follows.
In the MCQMC community, we focus largely on studying QMC and RQMC methods to estimate 
integrals that represent the mathematical expectations of certain random variables.
In applications, the output random variable $X$ of interest often represents a random 
cost or performance measure.  
But why estimate only the mean (the expectation) $\EE[X]$?
Data from simulation experiments can provide much more useful information than just an
estimator and a confidence interval for $\EE[X]$.  
When the number $n$ of realizations of $X$ is large enough, it permits one to estimate 
the entire distribution of $X$.  And when $X$ is a continuous random variable, 
this distribution is best visualized by showing its density.
On the other hand, density estimation from a sample of $n$ independent realizations of $X$
is known to be a difficult problem in statistics.
The leading density estimation methods, e.g., kernel density estimators (KDEs), 
only achieve a convergence rate of $\cO(n^{-4/5})$ for the mean square error (MSE) on the density
at a given point, compared to a $\cO(n^{-1})$ rate for the expectation \new{with MC}.
The main question raised in our 2015 discussion was:
We know that RQMC can improve the $\cO(n^{-1})$ rate for the mean,
but can it also improve the $\cO(n^{-4/5})$ rate for the density, by how much, and how?

Of course, this question makes sense only when the samples of $X$ are obtained by simulation
from a model, and not in the situation where $n$ independent observations of $X$ are given a priori.
When the observations are generated from a model, there is room to change the way we generate them
and construct the estimator, and in particular we may use RQMC points in place of independent
uniform random numbers to generate the observations of $X$.
Following this discussion, PL started exploring empirically what happens when we do this with an ordinary KDE.
That is, what happens with the variance and MSE of the KDE estimator when the $n$ observations of $X$
are generated by simulation using a set of $n$ RQMC points in place of $n$ independent points,
just like we do when estimating the mean.
After much experiments and theoretical work with co-authors, this led to \cite{vBEN21a}.
In that paper, we were able to prove an upper bound for the MSE with KDE+RQMC, but 
this bound converges at a faster rate than $\cO(n^{-4/5})$ only when the dimension $s$ is very small.
For moderate and large $s$, the bound converges at a slower rate than for crude Monte Carlo (MC),
although the observed MSE was never larger than for MC in our experiments.  
The reason for the slow rate for the bound is that when increasing $n$, 
we need to reduce the bandwidth of the 
KDE to reduce the square bias and the MSE, but reducing the bandwidth increases rapidly the variation 
of the estimator as a function of the uniform random numbers, and this hurts the RQMC estimator.

We understood that for RQMC to be effective, we need smoother density estimators.
In January 2017, while PL was visiting A.~Owen at Stanford University to work on \cite{vBEN21a}
he attended a talk by S.~Asmussen who (by pure coincidence) was presenting \cite{vASM18a}, 
in which he shows how to obtain an unbiased density estimator for a sum of independent random 
variables by conditional Monte Carlo. The conditioning is done by hiding the last variable in the sum and
taking the density of the last variable right-shifted by the sum of other variables as a density estimator.
We extended this idea to more general simulation models and this gave us 
what we needed to obtain smooth unbiased and RQMC-friendly density estimators. 
This led to the conditional density estimator (CDE) studied in \cite{vLEC19a},
also presented in 2018 at a SAMSI workshop on QMC methods
in North Carolina and at a RICAM workshop in Austria. 
The idea of this CDE method is to define a continuous estimator of the 
CDF $F(x)$ by conditioning, 
and take its sample derivative with respect to $x$ as a density estimator.
Under appropriate conditions, this provides an unbiased density estimator,
and when further favorable conditions hold, this estimator can be smooth and RQMC-friendly.
In March 2021, while we were finalizing this paper, Mike Fu pointed out that 
\cite{oFU06b} already contains an example in which he uses conditional Monte Carlo to estimate the 
density of the length of the longest path in a six-link network in which the last link is shared 
by all paths.  His unbiased density estimator is essentially the same as in \cite{vASM18a}: 
it is the density of the length of the last link, right-shifted by the length of the longest path 
up to that link.

At the Eleventh International Conference on Monte Carlo Methods and Applications (MCM), 
in July 2017, the authors of \cite{vLAU19a} presented a different approach that can provide
an unbiased density estimator for a sum of random variables as in \cite{vASM18a},
except that the variables can be dependent.
This approach can be generalized to obtain a continuous CDF estimator and then an unbiased 
density estimator, via the likelihood ratio (LR) simulation-based derivative estimation 
method \cite{oGLY87a,oLEC90a} and a clever change of variable, and by taking again
the sample derivative of this CDF estimator.  
This likelihood ratio density estimator (LRDE) is discussed in Section~\ref{sec:lr-rqmc} 
and also in \cite{vLEC21a}.  We also explain how it can be combined with RQMC.

A generalized version of the LR gradient estimator method, named GLR,
was proposed in \cite{oPEN18a} to handle situations in which neither the usual LR estimator 
nor the direct sample derivative apply, because of discontinuities.
In \cite{vLEI18a}, the authors sketch out how this GLR method could be used to 
obtain an unbiased density estimator.
Their general formulas are not easy to understand and implement, but more convenient 
formulas for these GLR density estimators are given in Theorem 1 of \cite{oPEN20a}. 
A modified version of the GLR named GLR-U was developed recently in \cite{oPEN21a}
to handle large classes of situations that could not be handled easily by the original 
GLR from \cite{oPEN18a}. The model of \cite{oPEN21a} is expressed explicitly 
in terms of independent uniform random variables over $(0,1)$.  
Density estimators can also be obtained by this method.

All these LR and GLR methods use a multivariate change of variable of some sort.  
They provide unbiased density estimators that are often not smooth with respect to the 
underlying uniforms, so their direct combination with RQMC does not always bring much gain.
However, it is often possible to smooth out the LR, GLR, or GLR-U density estimator
by conditioning just before applying RQMC.

The aim of this paper is to provide an overview of these recent developments on
density estimation for simulation models, by MC and RQMC.
We summarize the main theoretical results and give numerical illustrations 
on how the estimators behave, using simple examples.

The remainder is organized as follows.
In Sections~\ref{sec:density} and \ref{sec:rqmc}, we recall basic facts about 
one-dimensional density estimation and RQMC sampling.
In Section~\ref{sec:kde-rqmc}, we summarize what happens when we directly combine a KDE with RQMC. 
In Section~\ref{sec:cde-rqmc}, we discuss the CDE and its combination with RQMC.
In Section~\ref{sec:lr-rqmc}, we examine the LR and GLR density estimators.
Section~\ref{sec:examples} gives numerical illustrations. 
We wrap up with a conclusion in Section~\ref{sec:conclusion}.

\section{Basic density estimation}
\label{sec:density}

Let $X$ be a continuous real-valued random variable with CDF $F$ and density $f$.
The goal is to estimate the density $f$ over a finite interval $[a,b]$,
from a sample $X_1,\dots,X_n$ of $n$ realizations of $X$ (not necessarily independent).
This problem has been studied at length in statistics for the case where 
$X_1,\dots,X_n$ are independent \cite{tSCO15a}.
To measure the quality of an arbitrary density estimator $\hat f_n$ based on this sample,
we will use the \emph{mean integrated square error} (MISE), which is the integral of the 
MSE over the interval $[a,b]$:
\begin{equation}
\label{eq:mise}
  \MISE = \MISE(\hat f_n) = \int_a^b \EE [\hat f_n(x) - f(x)]^2 \d x = \IV + \ISB 
\end{equation}
where  
\[
  \IV = \int_a^b \EE (\hat f_n(x) - \EE [\hat f_n(x)])^2 \d x  \quad\mbox{ and }\quad
	\ISB = \int_a^b (\EE [\hat f_n(x)] - f(x))^2 \d x
\]
are the \emph{integrated variance} (IV) and the \emph{integrated square bias} (ISB), respectively.

Two popular types of density estimators are histograms and KDEs.
To define a \emph{histogram}, one can partition $[a,b]$ into ${m}$ intervals of length
${h}=(b-a)/m$ and put
\[
  \hat{f}_{n}(x) = \frac{n_{j}}{nh} \ \mbox{ for } x \in I_{j}=[a+(j-1)h, a+jh), \ \ j=1,...,m,
\]
where ${n_{j}}$ is the number of observations $X_i$ that fall in interval $I_j$.
To define a KDE \cite{tPAR62a,tSCO15a}, select a \emph{kernel} ${k}$ 
(usually a unimodal symmetric density centered at 0) 
and a \emph{bandwidth} ${h} > 0$ (an horizontal stretching factor for the kernel), and put
\[
 \hat{f}_{n}(x) = \frac{1}{nh} \sum _{i=1} ^{n}  k\left(\frac{x-X_{i}}{h}\right).
   \label{eq:kde}
\]
These two density estimators are biased.  
Asymptotically, when $n\to\infty$ and $h\to 0$ jointly, 
in the case of independent samples $X_1,\dots,X_n$, the IV and ISB behave as
\[
  \MISE ~=~  {\IV} + {\ISB} ~\sim~ {{C}}/{(nh)} + B h^{\alpha}
\]
where $C$, $B$, and $\alpha$ depend on the method.
The asymptotically optimal $h$ is then
\[
  {h^*} = \left({C}/{(B\alpha n)}\right)^{1/(\alpha+1)}
\]
and it gives $\MISE \sim K n^{-\alpha/(1+\alpha)}$ for some constant $K$. 
Table~\ref{tab:mise-rates} gives expressions for $C$, $B$, $\alpha$, $h^*$, and $\alpha/(1+\alpha)$,
for histograms and KDEs, with independent samples.
It uses the following definitions, for any $g : \RR\to\RR$: 
\begin{eqnarray*}
  {R(g)} = \int_a^b (g(x))^2 \d x &\mbox{ and }&
  {\mu_{r}(g)} = \int_{-\infty}^\infty x^r g(x) \d x \quad\mbox{ for } r = 0 \mbox{ and } 2.
\end{eqnarray*}

\begin{table}[hbt]
\caption{Constants involved in the convergence rates of the MISE for histograms and KDEs}
\label{tab:mise-rates}
\begin{center}
\setlength{\tabcolsep}{8pt}
		\begin{tabular}{r|ccccccc}
			\hline
			          &  $C$  &  $B$ &  $\alpha$  & $h^*$ & MISE  \\			
			\hline			
			Histogram &   1   &  ${R(f')}/{12}$  &  2 
			          & $(n R(f')/6)^{-1/3}$ & $\cO(n^{-2/3})$ \\[8pt]
			KDE   	  & $\mu_0(k^2)$ & ${(\mu_2(k))^2 {R(f'')}}/{4}$ & 4 
			          & $\left(\displaystyle\frac{\mu_0(k^2)}{(\mu_2(k))^2 R(f'') n}\right)^{1/5}$ 
								& {$\cO(n^{-4/5})$} \\	
			\hline
\end{tabular} 	
\end{center}
\end{table}

Note that these expressions hold under the simplifying assumption that $h$ must be the same 
all over $[a,b]$. One may often do better by varying the bandwidth over $[a,b]$,
but this is more complicated.
To estimate $h^*$ in practice, one can estimate $R(f')$ and $R(f'')$ by using a KDE to estimate
$f'$ and $f''$ (very roughly).  This type of crude (plugin) estimate is often good enough.
In the following, we will see how to improve on these MISE rates and values in a simulation setting,
\new{by reducing the variance.  In general, using RQMC points instead of MC does not change the bias.}

\section{RQMC}
\label{sec:rqmc}

We recall here some basic principles of RQMC used in the forthcoming sections.
For more extensive coverages, see \cite{rDIC10a,vLEC09f,vLEC18a,rNIE92b}, for example.
Suppose we want to estimate $\mu = \EE[g(\bU)]$ where $\bU = (U_1,\dots,U_{s})$ has the 
uniform distribution over the $s$-dimensional unit cube $(0,1)^s$ and $g : (0,1)^s \to \RR$.  
With standard MC, we draw $n$ independent random points $\bU_i$ uniformly over $(0,1)^s$ 
and we estimate the expectation by the average 
\begin{equation}
\label{eq:aver}
  \hat\mu_{n,\mc} = \frac{1}{n} \sum_{i=1}^{n} g(\bU_i).
\end{equation}
With RQMC, we replace the independent random points $\bU_i$ by a set 
of \emph{dependent} random points 
$\tilde P_n =\{\bU_1,\dots,\bU_{n}\} \subset (0,1)^{s}$ such that 
(1) the point set $\tilde P_n$ covers the unit hypercube very evenly 
(in a sense that must be precisely defined) with probability 1; and
(2) each point $\bU_i$ has the uniform distribution over $(0,1)^s$.
Then we estimate the expectation by the same average as in (\ref{eq:aver}), 
which we now denote $\hat\mu_{n,\rqmc}$.
For various spaces ${\cH}$ of functions $g$, usually Hilbert or Banach spaces, 
we have inequalities of the form
\begin{equation}
\label{eq:kh}
  \Var[\hat\mu_{n,\rqmc}] \le \cD^2(P_n) \cdot \cV^2(g)
\end{equation}
where $\cD(P_n)$ measures the \emph{discrepancy} of $P_n$ (with respect to the uniform distribution)
and $\cV(g)$ measures the variation of the function $g$.  
For many of these function spaces, we also know explicitly how to construct RQMC point sets 
for which $\cD(P_n) = \cO(n^{-\alpha/2} (\log n)^{s-1})$ for some ${\alpha} > 1$ 
\cite{rDIC10a,rGOD19a,vLEC16a}.
\hflorian{Perhaps call this $\beta$, so we do not confuse it with the parameter for the bias in the KDE.}
This leads to 
\[
  \Var[\hat\mu_{n,\rqmc}] = \cO(n^{-\alpha} (\log n)^{2(s-1)})
\]
when $\cV(f) < \infty$. A classical case is the standard Koksma-Hlawka inequality, 
for which $\alpha=2$, $\cD(P_n) = \cD^*(P_n)$ is the star discrepancy,
and $\cV(g)$ is the variation in the sense of Hardy and Krause, defined by
\begin{eqnarray}
\label{eq:hardy-krause}
  \cV(g) = \cV_{\rm HK}(g) = \sum_{\emptyset\not=\frakv\subseteq\{1,\dots,s\}} \int_{(0,1)^{|\frakv|}} 
	   \left|\frac{\partial^{|\frakv|}}{\partial\bu_{\frakv}}g(\bu_{\frakv},\bone) \right| \d\bu_{\frakv},
\end{eqnarray}
where $\bu_{\frakv}$ is the vector of coordinates whose indices belong to $\frakv$,
\new{$|\frakv|$ is the cardinality of $\frakv$,} 
and under the assumption that this expression is well defined.
\hflorian{This formula only holds under some smoothness assumptions.}%
The main construction methods for $P_n$ are lattice rules
and digital nets.

In the context of density estimation, the average in (\ref{eq:aver}) is replaced 
by the density estimator $\hat f_n(x)$ at a given point $x$.
If our density estimator can be written as an average of the form 
\begin{equation}
\label{eq:density-aver}
  \hat f_n(x) = \frac{1}{n} \sum_{i=1}^{n} \tilde g(x,\bU_i)
\end{equation}
where $\tilde g$ is a sufficiently smooth function of its second argument, 
then we can apply the RQMC theory just described to this density estimator by replacing 
the function $g(\cdot)$ by $\tilde g(x,\cdot)$.
We look at this in the next few sections.

\section{Kernel density estimators with RQMC}
\label{sec:kde-rqmc}

The KDE at a given point $x \in [a,b]$ is 
\[
 \hat f_n(x) = \frac{1}{n} \sum_{i=1}^n    
               \frac{1}{h} k\left(\frac{x-g(\bU_i)}{h}\right)    
              = \frac{1}{n} \sum_{i=1}^n  \tilde g(x,\bU_i).
\]
\new{We assume that the kernel $k$ is a smooth probability density, symmetric about 0,
and at least $s$ times differentiable everywhere.}
With RQMC points $\bU_i$, this is an RQMC estimator of 
$\EE[\tilde g(x,\bU)] = \EE[\hat f_n(x)]$.
RQMC does not change the bias of this density estimator, but it may reduce
$\Var[\hat f_n(x)]$, which would reduce in turn the IV and the MISE.

To prove RQMC variance bounds via (\ref{eq:kh}), we need to bound the variation $\cV(\tilde g)$.
This was done in \cite{vBEN21a} for the classical Hardy-Krause variation (\ref{eq:hardy-krause}),
which is bounded if and only if all the partial derivatives
\[
   \frac{\partial^{|\frakv|}}{\partial\bu_{\frakv}} \tilde g(x,\bu)
 = \frac{1}{h} \frac{\partial^{|\frakv|}}{\partial\bu_{\frakv}} k\left(\frac{x-g(\bu)}{h}\right)
\]
exist and are uniformly bounded.  The derivatives with respect to $k$ are easily bounded
for instance if $k$ is a standard normal \new{density (the Gaussian kernel).}
However, when expanding the derivatives via the chain rule, we obtain terms in $h^{-j}$ 
for $j=2,\dots,|\frakv|+1$.  \new{The dominant term asymptotically is the term for $|\frakv| = s$,
and it grows in general as}
$h^{-s-1} \left| k^{(s)}\left((x-g(\bu))/h\right) \prod_{j=1}^s g_{\{j\}}(\bu)\right| 
= \cO(h^{-s-1})$ when $h\to 0$, 
\new{where $g_{\{j\}}$ is the derivative of $g$ with respect to its $j$th coordinate.}
We can bring it down to $\cO(h^{-s})$ via a change of variables, which leads to the following 
result proved in \cite{vBEN21a}:

\begin{proposition}
Let $g:[0,1]^{s}\to\RR$ be \emph{piecewise monotone} in each coordinate $u_j$ when 
the other coordinates are fixed.
Assume that all first-order partial derivatives of $g$ are continuous and that
$\|g_{\mathfrak{w_1}} g_{\mathfrak{w_2}}\cdots g_{\mathfrak{w_\ell}}\|_1 < \infty$
for all selections of non-empty, mutually disjoint index sets 
$\mathfrak{w_1},\dots,\mathfrak{w_\ell}\subseteq \{1,\dots,s\}$,
\new{where $g_{\mathfrak{w}}$ is the derivative of $g$ with respect to all the coordinates
in the index set $\mathfrak{w}$.}

Then the Hardy-Krause variation of  \new{$\tilde{g}(x,\cdot)$ for any fixed $x\in [a,b]$} satisfies
\[
 \cV_{\rm HK}(\tilde{g}(x,\cdot)) \leq {c} h^{-s} + \cO(h^{-s+1})
\]
\new{for some constant $c > 0$ given in \cite{vBEN21a}},
and with RQMC point sets having \new{a star discrepancy} 
$\cD^{*}(P_n)=\cO(n^{-1+\epsilon})$ for all $\epsilon > 0$ when $n\to\infty$, we obtain
\[
  \IV = \cO(n^{-2+{\epsilon}} h^{-2s})  \quad\mbox{ for all }  {\epsilon} > 0.  
\]
\new{RQMC does not change the bias, so the ISB has exactly the same expression as for MC.}
By picking $h$ to minimize the MISE bound, we get $\MISE = \cO(n^{-4/(2+s) + \epsilon})$.  
\end{proposition}

This rate for the MISE is worse than the MC rate when $s \ge 4$.   
The factor $h^{-2s}$ in the IV bound really hurts.  
On the other hand, this is only an upper bound, not the actual IV.
Proposition 4.4 of \cite{vBEN21a} also shows via a different analysis that for the KDE, 
there exist RQMC constructions for which the asymptotic decrease rate of the IV is not 
worse than for MC.

\hpierre{--- Perhaps add KDE + stratification? }

\section{Conditional density estimation with RQMC}
\label{sec:cde-rqmc}

To estimate the density $f(x) = F'(x)$, one may think of simply taking the sample derivative
of an unbiased estimator of the CDF $F(x)$. The simplest unbiased estimator of this CDF is 
the \emph{empirical CDF} 
\[
  \hat F_n(x) = \frac{1}{n} \sum_{i=1}^n \II[X_i\le x].
\]
However $\d {\hat F_n(x)} /\d x = 0$ almost everywhere, 
so this \emph{cannot} be a useful density estimator!
We need a smoother CDF estimator, which should be at least continuous in $x$.

One effective way of smoothing an estimator and often make it continuous
is to replace it by its conditional expectation given partial (filtered) information.
This is \emph{conditional Monte Carlo} (CMC) \cite{sASM07a}.
That is, one replaces the indicator $\II[X_i\le x]$ in the expression of $\hat F_n(x)$ above
by the conditional CDF $F(x \mid \cG) = \PP[X_i \leq x \mid \cG]$,
where ${\cG}$ is a sigma-field that contains not enough information to reveal $X$ 
but enough to compute $F(x \mid \cG)$, then one takes the sample derivative.
We call it the \emph{conditional density estimator} (CDE).
\new{For more details about the CMC method in general and the choice of $\cG$ in specific cases,
see for example \cite{sASM07a,oFU97a,oLEC94a}.  For examples in the context of density estimation,
see \cite{vLEC19a} and the examples in Section~\ref{sec:examples}.
We assume here that we can compute the conditional density either directly or numerically by an 
iterative algorithm.}
The following proposition, proved in \cite{vLEC19a},
gives sufficient conditions for this CDE to be an unbiased density estimator with finite variance.
 
\begin{proposition}
\label{prop:cde}
Suppose that for all realizations of $\cG$, $F(x \mid \cG)$ is a \emph{continuous} function 
of $x$ over $[a,b]$, \emph{differentiable} except perhaps over a denumerable set of points 
$D(\cG) \subset [a,b]$, and for which $f(x \mid \cG) = F'(x \mid \cG) = \d F(x \mid \cG) /\d x$ 
(when it exists) is bounded uniformly in $x$ by a random variable $\Gamma$ 
such that $\EE[\Gamma^2] \le K_\gamma < \infty$.
Then, for all $x\in[a,b]$, $\EE[f(x \mid \cG)] = f(x)$ and $\Var[f(x \mid \cG)] < K_\gamma$.
Moreover, if $\cG \subset \tilde\cG$ both satisfy the assumptions of this proposition, then 
$\Var[f(x \mid \cG)] \le \Var[f(x \mid \tilde\cG)]$.
\end{proposition}

For a sample of size $n$, the CDE becomes 
\[
  \hat f_{{\rm cde},n}(x) = \frac{1}{n} \sum_{i=1}^n f(x \mid \cG^{(i)})
\]	
where $\cG^{(1)},\dots,\cG^{(n)}$ are $n$ ``realizations'' of $\cG$.
When the $n$ realizations are independent we have 
$\Var[\hat f_{{\rm cde},n}(x)] \le K_\gamma / n = \cO(n^{-1})$.

To combine the CDE with RQMC, we want to write $f(x \mid \cG) = \tilde g(x,\bu)$ 
for some function $\tilde g : [a,b]\times [0,1)^s \to\RR$.
This function $\tilde g(x,\cdot)$ will be used in (\ref{eq:density-aver}).
The combined CDE+RQMC estimator is then defined by
\begin{equation}
\label{eq:hatf-cde-rqmc}
  \hat f_{\cde-\rqmc,n}(x) = \frac{1}{n} \sum_{i=1}^n \tilde g(x,\bU_i).
\end{equation}
where $\{\bU_1,\dots,\bU_n\}$ is an RQMC point set. 

If $\tilde g(x,\cdot)$ has \emph{bounded variation}, 
then we can get an \emph{$\cO(n^{-2+\epsilon})$ rate for the MISE},
and sometimes better.   This holds in several examples that we tried.
If $\tilde g(x,\cdot)$ has \emph{unbounded variation}, RQMC may still reduce the IV,
but there is no guarantee.

\section{Likelihood ratio density estimators}
\label{sec:lr-rqmc}

\hflorian{I replaced several $d$'s by $s$'s in this section and Section \ref{sec:glr-rqmc}, to unify 
  notation. Perhaps briefly check if I have missed some.}
There are situations where a CDE as in Section~\ref{sec:cde-rqmc} might be too difficult to obtain.
An alternative can be a \emph{likelihood ratio density estimator} (LRDE), defined as follows.
Suppose that ${X = h(\bY)}$ where $\bY$ has known density ${f_{\bY}}$ over $\RR^s$,
and we know how to generate it and compute $X = h(\bY)$.
For simplicity, let $x > 0$ (in case we are really interested in some $x\le 0$, 
we can simply add a constant to the function $h$).
We have
\[
  F(x) = \PP[h(\bY)\le x] 
	     = \int_{\RR^s} \II[h(\by)/x \le 1] f_{\bY}(\by) \d\by.
\]
\hflorian{We probably need to assume $x\geq 0$ to make this work. 
 You could just write that we assume that for simplicity.}%
We want to change this integrand into a continuous function of $x$, so we can take the 
derivative with respect to $x$ inside the integral.
One way to do this is to make a change of variable $\by \mapsto \bz = \bz(x)$ of the form 
$\by = \varphi_x(\bz)$, with Jacobian $|J_x(\bz)|$,
so that $\tilde h(\bz) = h(\varphi_x(\bz))/x$ no longer depends on $x$ for any given $\bz$.
We can then rewrite
\[
 F(x) = \int_{\RR^s} \II[\tilde h(\bz)\le 1] 
        f_{\bY}(\varphi_x(\bz)) |J_x(\bz)| \d\bz.
\]
In a small open neighborhood of a given $x_0 \in [a,b]$, we have
\[
 F(x) = \int_{\RR^s} \II[\tilde h(\bz)\le 1] L(\bz;x,x_0)
         f_{\bY}(\varphi_{x_0}(\bz)) |J_{x_0}(\bz)|\d\bz
\]
where 
\[
 L(\bz;x,x_0)
  = \frac{f_{\bY}(\varphi_x(\bz)) |J_x(\bz)|}{f_{\bY}(\varphi_{x_0}(\bz)) |J_{x_0}(\bz)|}
\]
is the \emph{likelihood ratio} between the density of $\bz$ at $x$ and at $x_0$.
Under appropriate conditions:
\begin{eqnarray*}
 f(x) &=& \frac{\d}{\d x} \int_{\RR^s} \II[\tilde h(\bz)\le 1] 
           L(\bz;x,x_0) f_{\bY}(\varphi_{x_0}(\bz)) |J_{x_0}(\bz)|\d\bz \\
       &=& \int_{\RR^d} \II[\tilde h(\bz)\le 1] 
           {\left(\frac{\d}{\d x} L(\bz;x,x_0)\right)} 
					 \frac{f_{\bY}(\varphi_x(\bz)) |J_x(\bz)|}{L(\bz;x,x_0)} \d\bz \\
       &=& \int_{\RR^d} \II[\tilde h(\bz)\le 1] 
           {\left(\frac{\d}{\d x} \ln L(\bz;x,x_0)\right)} 
					  f_{\bY}(\varphi_x(\bz)) |J_x(\bz)| \d\bz \\
       &=& \int_{\RR^s} \II[h(\by)\le x] S(\by,x) f_{\bY}(\by) \d\by
\end{eqnarray*}
where 
\begin{eqnarray*}
  S(\by,x) &=& 
  \frac{\d \ln L(\bz;x,x_0)}{\d x} 
	 = (\nabla(\ln f_{\bY})(\by)) \cdot (\nabla_x \varphi_x(\bz))  +  
	     \frac{\d \ln |J_x(\bz)|}{\d x} 
\end{eqnarray*}
is the \emph{score function} associated with $L$.
\hpierre{I think if we still want to submit the separate paper on LRDE \cite{vLEC21a}, 
 then we should cut out much of the derivation details around here and refer to the other paper.}%
This gives the unbiased LRDE
\begin{equation}
\label{eq:lrde}
  \hat f_{\lrde}(x) = \II[h(\bY)\le x] \, S(\bY,x)
\end{equation}
where $\bY \sim f_{\bY}$.  
Here, $\bY$ can have a multivariate distribution for which conditioning is hard
whereas $S(\bY,x)$ may be easier to compute.

This LR approach has been widely used to estimate the derivative of $\EE[h(\bY)]$
with respect to a parameter of the distribution of $\bY$ 
\cite{sASM07a,oGLY87a,oGLY95a,oLEC90a}.
Laub et al. \cite{vLAU19a} obtained (via a different argument) the estimator (\ref{eq:lrde}) 
for the special case where $h(\bY)$ is a sum of random variables.
The following is proved in \cite{vLEC21a}.

\begin{proposition}
\label{prop:lr}
Suppose that with probability one over realizations of $\bY = \varphi_x(\bZ)$,
$f_{\bY}(\varphi_x(\bZ)) |J_x(\bZ)|$ is continuous in $x$ over $[a,b]$
and is differentiable in $x$ except perhaps at a countable set of points $D(\bY) \subset [a,b]$.
Suppose that there is also a random variable ${\Gamma}$ defined over the same probability space as $\bY$,
such that $\EE[\Gamma^2] < \infty$, and for which
\[
  \sup_{x\in [a,b]\setminus D(\bY)} \left|\II[h(\bY)\le x] \, S(\bY,x)\right| \le \Gamma.
\]
Then, $\hat f_{\lrde}(x) = \II[h(\bY)\le x] \,S(\bY,x)$ is an unbiased estimator 
of $f(x)$ at almost all $x\in[a,b]$, with variance bounded uniformly by $\EE[\Gamma^2]$.
\end{proposition}

Note that the unbiased LRDE in (\ref{eq:lrde}) is usually discontinuous in the underlying 
uniforms, because of the indicator function, so it is not a smooth RQMC-friendly estimator.
One can think of making it continuous by taking its conditional expectation.
On the other hand, when we can find a conditioning that makes the indicator continuous,
then we may be able to apply the CDE instead and this is usually more effective, according to
our experiments.  The LRDE is nevertheless useful for the situations in which a CDE is 
difficult to obtain.

\section{Generalized likelihood ratio estimators}
\label{sec:glr-rqmc}

Peng et al. \cite{oPEN18a} proposed a \emph{generalized likelihood ratio} (GLR) method
that generalizes the LR derivative estimation approach.
Peng et al. \cite{oPEN20a} gave an adaptation of this method to density estimation.
It goes as follows.  
Let ${X = h(\bY) = h(Y_1,\dots,Y_s)}$ for some random variables $Y_1,\dots,Y_s$, 
and assume that $X$ is a continuous random variable with (unknown) density $f$.
Let $A(x,\epsilon) = \{\by \in\RR^s : x-\epsilon \le h(\by) \le x+\epsilon\}$,
which is the inverse image of an $\epsilon$-neighborhood of $x$ by $h$.
Suppose there is an $\epsilon_0 > 0$ such that
\[
  \lim_{\epsilon\to 0} \sup_{x\in [a-\epsilon_0, b+\epsilon_0]} \lambda(A(x,\epsilon)) = 0,
\]
where $\lambda$ is the Lebesgue measure on $\RR^s$.
\hpierre{In their statement, this is really the Lebesgue measure, and not the measure with
  respect to the density $f_{\bY}$.  This appears quite restrictive.
	For example, in one dimension, if $Y$ is standard normal and $g(y) = 1-e^{-y}$, 
	the condition will not be satisfied over $[0,1]$. But this could probably work if we
	express $g$ as a function of a $U(0,1)$ instead. }%
Select some index $j \in\{1,\dots,s\}$ for which $Y_j$ is a continuous random 
variable with CDF ${F_j}$ and density ${f_j}$, and is independent of $\{Y_k,\, k\not=j\}$.
Let ${h_{(j)}(\by)} := \partial h(\by) /\partial y_j$,
${h_{(jj)}(\by)} := \partial^2 h(\by) /\partial y_j^2$, and 
\[
 \Psi_j(\by) = \frac{\partial \ln f_j(y_j) /\partial y_j - h_{(jj)}(\by) / h_{(j)}(\by)}{h_{(j)}(\by)},
\]
where all these derivatives are assumed to exist.
Suppose that there are functions $v_\ell : \RR\to\RR$ for $\ell=1,\dots,s$ such that
$|h_{(j)}(\by)|^{-1} \le \prod_{\ell=1}^d v_{\ell}(y_{\ell})$ and
\[
  \lim_{y \to\pm\infty} v_j(y) f_j(y) = 0 \mbox{ \ and \ }
  \EE[v_j(Y_j)] < \infty.
\]
\hpierre{Here, I replaced a product by a single $v_j$; we must check the proof to see if this is ok.}%
\hflorian{Your assumption is stronger. In your case $h_{(j)}^{-1}$ has to be bounded uniformly in all
variables but the $j$th one. Also, I see only one step in the proof where this is used, and it still
checks out.}%
Finally, suppose also that 
$\EE[\II[X \le x] \Psi_j^2(\bY)] < \infty$.
Under all these conditions, a simple modification of the proof of Theorem~1 in \cite{oPEN20a} yields
the following:

\begin{proposition}
\label{prop:glr}
With the assumptions just given, 
$D_j(x,\bY) = \II[X \le x] \Psi_j(\bY)$ is an unbiased and 
finite-variance estimator of the density $f(x)$ at $x$.
\end{proposition}

When the conditions hold for all $j=1,\dots,s$, as assumed in \cite{oPEN20a},
this gives $s$ unbiased estimators $D_1(x,\bY), \dots, D_s(x,\bY)$.
Instead of selecting only one of them, we can take a linear combination 
$D(x,\bY) = w_1 D_1(x,\bY) + \cdots + w_s D_s(x,\bY)$ where $w_1 + \cdots + w_s = 1$.
This is exactly equivalent to taking, say $D_1(x,\bY)$ as the base estimator and the 
$C_j = D_j(x,\bY) - D_1(x,\bY)$ as mean-zero control variates, for $j=2,\dots,s$,
because one has $D(x,\bY) = D_1(x,\bY) + w_2 C_2 \cdots + w_s C_s$.
Therefore, standard control variate theory \cite{sASM07a} can be used to optimize
the coefficients $w_j$.
When the conditions are satisfied only for certain values of $j$, then one can take the 
linear combination only for these values.
It may also happen that the assumptions are satisfied for no $j$,
in which case this method does not apply.

The GLR setting of \cite{oPEN18a} is more general.   
It permits one to estimate the derivative of $\EE[\varphi(g(\bY;\theta))]$ with respect to some 
parameter $\theta$, where $g(\cdot;\theta) : \RR^s\to\RR^s$ is continuous and 
one-to-one for the values of $\theta$ in the region of interest, so it corresponds 
to a multivariate change of variable in that region.
The authors provide a general form of the unbiased derivative estimator (see also \cite{oPEN21b}).
The general formula is rather complicated and it can be found in the papers.
One can use it in principle to estimate the density of $X$ by taking $\theta = x$ and 
selecting a $g$ for which $\varphi(g(\bY;\theta)) \equiv \II[X \le x] = \II[h(\bY)-x \le 0]$
and for which the assumptions of \cite{oPEN18a} are satisfied, when this is possible.
 
Peng et al.~\cite{oPEN21a} extended the range of applicability of GLR by developing GLR-U, 
a version of GLR in which the base model is expressed directly in terms of the 
underlying uniform random numbers.  That is, $\bY$ takes the form of a vector $\bU$ which 
has the uniform distribution over the unit hypercube $(0,1)^s$.
\hflorian{Perhaps ``replaced'' is too strong. What do you think? Maybe we could say 
``expressed'' again.}%
This new setting covers a larger class of models than in \cite{oPEN18a}, including situations
where the random variables are generated by inversion, by the rejection method, or via
Archimedean copulas, for example.
We outline how to use this method to estimate the density of $X$ over $[a,b]$.

The first step is to find a nonempty subset of the input variables $\Upsilon\subseteq\{1,\dots,s\}$,
which we will assume (without loss of generality) to be $\Upsilon = \{1,\dots,d\}$ for $1\le d\le s$,
together with a function $g(\cdot;x) = g_1(\cdot;x), \dots, g_d(\cdot;x)) : (0,1)^s\to\RR^d$ for which
$\varphi(g(\bU;x)) \equiv \II[X \le x]$ for all $x\in[a,b]$ and which satisfies the following assumptions.
For any $\bu\in(0,1)^s$, we decompose $\bu = (\bu^{(1)}, \bu^{(2)})$ where $\bu^{(1)}$ contains the 
first $d$ coordinates and $\bu^{(2)}$ the other ones.
When $\bu^{(2)}$ is fixed, $g(\cdot;x)$ becomes a function of $\bu^{(1)}$ only, which we denote
by $\tilde g(\cdot; \bu^{(2)},x)$.  An important condition is that this function $\tilde g$ must
be continuous and correspond to a multivariate change of variable, whose Jacobian $J_g(\bu; x)$
is a $d\times d$ invertible matrix whose element $(i,j)$ is $\partial g_i(\bu; x) /\partial u_j$.
For any $\bu = (u_1,\dots,u_s) \in (0,1)^s$ and $j=1,\dots,d$, 
let $\overline{\bu}_j$ and $\underline{\bu}_j$ be the vector $\bu$ 
\new{in the limit when $u_j\to 1$ from the left and the limit when $u_j\to 0$ from the right
(see \cite{oPEN21a,oPEN21b})}.
Define 
\[
  r_j(\bu; x) = -(J_g^{-1}(\bu; x))^\tr \cdot\be_j
\]
and 
\[
  v(\bu; x) = -\sum_{j=1}^d \be_j^\tr \cdot(J_g^{-1}(\bu; x)) 
	             \left(\frac{\d J_g(\bu; x)}{\d u_j} \right) (J_g^{-1}(\bu; x)) \cdot\bone
\]
where $\be_j$ is the $j$th unit vector, $\bone$ is a column vector of ones, 
and the derivative of $J_g(\bu; x)$ is element-wise. 
Then, under some mild regularity conditions, we have:
\begin{proposition}
The following is an unbiased density estimator at all $x\in[a,b]$:
\begin{equation}
\label{eq:glr-u}
 G(\bU,x) = \II[X \le x]\, \new{v(\bU;x)} 
            + \sum_{j=1}^s \left[\varphi(g(\overline{\bU}_j;x)) r_j(\overline{\bU}_j;x)
						                    -\varphi(g(\underline{\bU}_j;x)) r_j(\underline{\bU}_j;x)\right].
\end{equation}
\end{proposition}

Peng et al.~\cite{oPEN21a,oPEN21b} show how to apply this method in the special case where 
$X$ is the maximum of several variables, each one being the sum of certain $Y_j$'s
that are generated by inversion from the $U_j$'s. This may correspond to the length of the 
longest path between a source node to a destination node in a directed network, for example.
It works in the same way if the maximum is replaced by a minimum, and we will use it
in Section~\ref{sec:examples}.  The number $d$ of selected input variables in this case 
should be equal to the number of independent paths.

\section{Numerical illustrations}
\label{sec:examples}

We illustrate the applicability and performance of the various density estimators 
discussed here on a small shortest path example defined below. 
\hpierre{This is *not* a SAN model; we look at the shortest path.}%
We run the simulations with MC and RQMC.
For RQMC, we use Sobol' nets with direction numbers taken from \cite{iLEM04a}, 
and randomized by a left matrix scramble followed by a digital shift. 
Each RQMC experiment is repeated $m=100$ times independently. 
The performance is assessed via the estimated $\MISE$ for $n=2^{20}$ points. 
For RQMC, we also estimate the convergence rate as follows:
we assume that $\MISE \approx n^{-\beta}$ for some constant $\beta > 0$ and we estimate
$\beta$ by $\hat\beta$ using linear regression in log scale, 
based on observations obtained with $n=2^{13},2^{14},\dots,2^{20}$. 
\hflorian{Just checking: I used $\beta$ to distinguish it from the KDE parameter $\alpha$.
But we kept $\alpha$ for the variance rate. The MISE is of course sometimes not the same
as the variance, but should we use $\alpha$ here too to maintain the notion that it has
to do with convergence?}%
\hpierre{I think no, since $\alpha$ has a different meaning;  see Table 1.}
For MC, the rates are known theoretically to be $\beta=0.8$ for the KDE and $\beta=1$ 
for the other methods. For the experiments with the KDE, we select the bandwidth with
the same methodology as in \cite{vBEN21a}.
\hflorian{I think it is good to keep such technical details concentrated in one point.
 Since we basically have two experiments, I moved this part here and extended it by $\hat\beta$.}

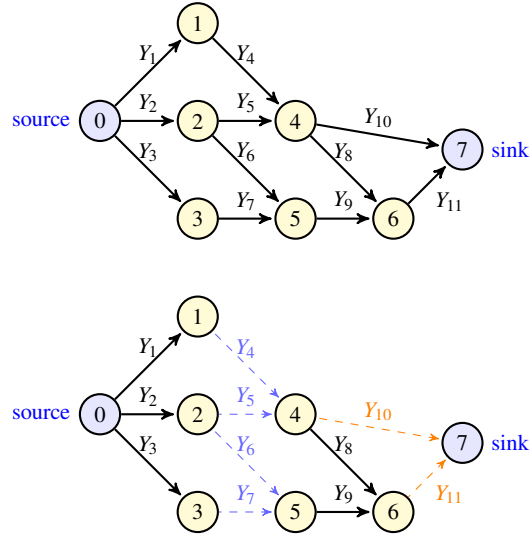
\begin{figure}[hbt]
\begin{center}
\small
\begin{tikzpicture}[node distance=1.3cm,>=stealth',auto]
\tikzstyle{node}=[circle,thick,draw=black,fill=yellow!20,minimum size=4mm]
\tikzstyle{nodeb}=[circle,thick,draw=black,fill=blue!10,minimum size=4mm]
    \node [nodeb] (n0) [label=left:{\small\color{blue} source}]  {0};
    \node [node] (n2) [right of=n0] {2}
          edge [pre,thick] node[above] {$Y_2$} (n0);
    \node [node] (n1) [above of=n2] {1}
          edge [pre,thick] node[above] {$Y_1$} (n0);
    \node [node] (n3) [below of=n2] {3}
          edge [pre,thick] node[above] {$Y_3$} (n0);
    \node [node] (n4) [right of=n2] {4}
          edge [pre,thick] node[above] {$Y_4$} (n1)
          edge [pre,thick] node[above] {$Y_5$} (n2);
    \node [node] (n5) [right of=n3] {5}
          edge [pre,thick] node[above] {$Y_6$} (n2)
          edge [pre,thick] node[above] {$Y_7$} (n3);
    \node [node] (n6) [right of=n5] {6}
          edge [pre,thick] node[above] {$Y_8$} (n4)
          edge [pre,thick] node[above] {$Y_9$} (n5);
    \node [nodeb] (n7) [above right of=n6, label=right:{\small\color{blue} sink}] {7}
          edge [pre,thick] node[above] {$Y_{10}$} (n4)
          edge [pre,thick] node[below right] {$Y_{11}$} (n6);
\end{tikzpicture}

\null\vspace{10pt}

\begin{tikzpicture}[node distance=1.3cm,>=stealth',auto,
    cuta/.style={color=blue!60,dashed},
    cutb/.style={color=red!50!yellow,dashed}
]
\tikzstyle{node}=[circle,thick,draw=black,fill=yellow!20,minimum size=4mm]
\tikzstyle{nodeb}=[circle,thick,draw=black,fill=blue!10,minimum size=4mm]
    \node [nodeb] (n0) [label=left:{\small\color{blue} source}]  {0};
    \node [node] (n2) [right of=n0] {2}
          edge [pre,thick] node[above] {$Y_2$} (n0);
    \node [node] (n1) [above of=n2] {1}
          edge [pre,thick] node[above] {$Y_1$} (n0);
    \node [node] (n3) [below of=n2] {3}
          edge [pre,thick] node[above] {$Y_3$} (n0);
    \node [node] (n4) [right of=n2] {4}
          edge [pre,cuta] node[above] {$Y_4$} (n1)
          edge [pre,cuta] node[above] {$Y_5$} (n2);
    \node [node] (n5) [right of=n3] {5}
          edge [pre,cuta] node[above] {$Y_6$} (n2)
          edge [pre,cuta] node[above] {$Y_7$} (n3);
    \node [node] (n6) [right of=n5] {6}
          edge [pre,thick] node[above] {$Y_8$} (n4)
          edge [pre,thick] node[above] {$Y_9$} (n5);
    \node [nodeb] (n7) [above right of=n6, label=right:{\small\color{blue} sink}] {7}
          edge [pre,cutb] node[above] {$Y_{10}$} (n4)
          edge [pre,cutb] node[below right] {$Y_{11}$} (n6);
\end{tikzpicture}
\end{center}
\caption{Upper panel: a directed network with 11 links.
   Lower panel: two selected minimal cuts \new{$\cL_1 = \{4,5,6,7\}$} (in light blue) 
	 and \new{$\cL_2 = \{10, 11\}$} (in orange) for this network.}
\label{fig:network}
\end{figure}

We consider an acyclic directed network as in Figure~\ref{fig:network},
with $s$ arcs.  For $j=1,\dots,s$, arc $j$ has random length $Y_j$
with continuous cdf $F_j$ and density $f_j$, and the $Y_j$ are assumed independent.
We generate $Y_j$ by inversion via $Y_j = F_j^{-1}(U_j)$ where $U_j\sim U(0,1)$.
We want to estimate the density of the length $X$ of the shortest path from the source
to the sink.

In the network of Figure~\ref{fig:network}, there are six different directed paths
from the source to the sink, each one being defined by a sequence of arcs. They are
$\cP_1=\{1,4,10\}$, $\cP_2=\{1,4,8,11\}$, $\cP_3=\{2,5,10\}$,
$\cP_4=\{2,5,8,11\}$, $\cP_5=\{2,6,9,11\}$, and $\cP_6=\{3,7,9,11\}$.
The length of path $p$ is $L_p = \sum_{j\in\cP_p} Y_j$ and 
the length of the shortest path is
\begin{equation}
\label{eq:shortest-path}
  X = h(\bY)  
	 = \min_{1\leq p \leq 6} L_p
	 = \min_{1\leq p \leq 6} \sum_{j\in\cP_p} F_j^{-1}(U_j).
\end{equation}

\begin{figure}[!htbp]
\centering
 \begin{tikzpicture} 
    \begin{axis}[ 
      title ={},
      width=.45\columnwidth,
      height=.3\columnwidth,
      xlabel=x,
      xmin=128.8,
      xmax=171.2,
	  grid,
	  scaled ticks=false,
	  tick label style={/pgf/number format/fixed}
      ] 
      \addplot[blue!85!black,line width=0.9pt] table[x=x,y=y] { 
      x  y 
      128.85385271574776  0.005457318266565828 
      129.78305568772825  0.006431528956597459 
      130.62709487060732  0.007418790935328492 
      131.6941590698811  0.008809569613529637 
      132.2859710842916  0.009649815553063792 
      133.26615964447737  0.01114776298037835 
      134.09184826220584  0.012508669646515761 
      134.88675738888176  0.013898852594707509 
      135.64165908600788  0.015285228454238823 
      136.7524813485565  0.017425577894496187 
      137.52403545521372  0.018968094734037926 
      138.30186717568813  0.020555263737978148 
      139.11436378407936  0.02223204213369328 
      139.9259202951322  0.023908362561864568 
      140.93066670453612  0.025956627349034825 
      141.9033727093892  0.02787648372733725 
      142.4947653759977  0.028997622155084703 
      143.23048150677516  0.030328678058546117 
      144.47376423050252  0.03237608046744566 
      144.97284205273147  0.03311301569898235 
      146.07950162177167  0.03454412169056291 
      146.68496951297152  0.03519680987256232 
      147.49012871038792  0.03590915581491165 
      148.39631468366306  0.03648548708288458 
      149.43982625785748  0.036838856907516114 
      150.08155742793622  0.03688695033381464 
      151.07042386745107  0.03670729718401574 
      151.78342082137976  0.036389062069026624 
      152.80731104460932  0.03566471538189319 
      153.7181813245336  0.034769514458091226 
      154.61678031358548  0.03367350601519015 
      155.46071419735546  0.03247086321565935 
      156.29364731623573  0.031139516019920554 
      157.0919667316473  0.029748660483549155 
      157.84564416090353  0.02835049774462761 
      158.51367269612754  0.027056164596808457 
      159.3456685246795  0.025390599811223526 
      160.57178622822732  0.022872954623194733 
      161.43856925033367  0.021081102040719838 
      162.07500497867935  0.01977437492056093 
      162.9689688652031  0.017969086972939502 
      163.8535372340157  0.016235444688048176 
      164.41685404629857  0.015166644846651289 
      165.66324578554816  0.012920016713494792 
      166.277925659001  0.01187956778487431 
      167.1936563480926  0.010420215102835176 
      168.11438240090934  0.00906777955075281 
      168.93829787940112  0.007957979163594354 
      169.56164675802702  0.007181752251750933 
      170.58003389915987  0.006029814987204856 
 }; 
%
    \end{axis}
  \end{tikzpicture}
\caption{Estimated density for the shortest path example.}
\label{fig:density}
\end{figure}
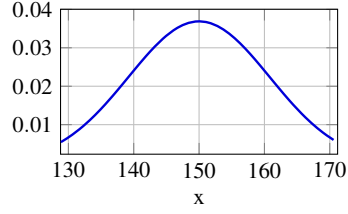

For our experiments, we assume that $Y_j$ is normal with mean $\mu_j=10 j$ 
and standard deviation $\sigma_j=j$ (to make things simple).  
The probability of negative arc lengths is negligible.
\new{(To be mathematically cleaner, we can truncate the normal density to $[0,\infty)$,
but it makes no visible difference in the numerical results.)}
We estimate the density of $X$ over $[a,b] = [128.8, 171.2]$, which covers 
about 95\% of the density.  This density is shown in Figure~\ref{fig:density}.
It is close to a normal distribution, which is not surprising because all the $Y_j$ are normal.

For the CDE, we select a directed minimal cut $\cL$ between 
the source and the sink, and we condition on $\cG = \{Y_j,\, j\not\in\cL\}$, similarly as for
the SAN example in \cite{vLEC19a}.  
If $P_j + Y_j$ is the length of the shortest path that goes through arc $j$ for $j\in\cL$,
then conditional on $\cG$, each $P_j$ is known and the conditional cdf of $X$ is
\begin{equation}
\label{eq:cmc-estimator}
  F(x\mid\cG) = \PP\left[X \le x \mid \{P_j : j\in {\cL}\}\right] 
							= 1 - \prod_{j\in\cL} (1-F_j(x - P_j)).
\end{equation}
If the $Y_j$'s for $j\in\cL$ are continuous variables,  
then the conditional density
\[
  f(x\mid \cG) = \frac{\d}{\d x} F(x\mid \cG)  
	 = \sum_{j\in\cL} f_j(x - P_j)  \prod_{l\in\cL,\, l\not=j} (1-F_l(x - \new{P_l}))
\]
is an unbiased density estimator.   
In our numerical experiments, we try the two cuts $\cL_1$ and $\cL_2$ shown on 
the lower panel of Figure~\ref{fig:network}.

For the LRDE we notice that $h(\bY)$ is the minimum over the  lengths of six possible paths. 
These lengths, in turn, are simple sums of several of the $Y_j$, so we have $h(c\bY) = c h(\bY)$
for any constant $c > 0$.  Therefore, with the change of variables 
$\varphi_x(\bz) = x\,\bz$ one obtains that $h(\varphi_x(\bz))/x = h(\varphi_x(\bz)/x)= h(\bz)$ 
is independent of $x$.  For $x > 0$ this leads to the LRDE 
\begin{equation}
\label{eq:lrde_paths}
 \hat{f}_{\lrde}(x) 
  = \II[h(\bY)\leq x]\,\, x^{-1}\left( -\sum_{j=1}^{s}(Y_j-\mu_j) Y_j \sigma_{j}^{-2} + s \right).
\end{equation}

For GLR, the estimator in Proposition~\ref{prop:glr} does not apply to this example,
because for the function $h$ given in \eqref{eq:shortest-path}, for any choice of $j$,
the required derivatives do not always exist.
For the GLR-U, we want to find a subset of indices and a function $g$ that satisfy 
the required conditions. In particular, $\tilde g$ must be a one-to-one continuous map 
between the selected inputs $U_j$ and a selected subset of the path lengths $L_p$, 
so that the latter subset is sufficient to determine $X$
and the Jacobian $J_g(\cdot; x)$ of this mapping is invertible.
Note that the six path lengths are not independent: we have $L_1 + L_4 = L_2 + L_3$. 
But after removing one of these four paths, there is no linear relationship between any of
the five $L_p$'s that remain.  Then we must select five input variables $U_j$ for which 
the mapping $g$ between those selected $U_j$'s and the five $L_p$'s is one-to-one when the other 
$U_j$'s are fixed.  There are several possibilities for the selection of these five indexes $j$
for the inputs, each one leading to a different estimator.
We will try two of them in our experiments, namely 
$\cJ_1 = \{1,2,3,6,8\}$ and $\cJ_2 = \{5,7,8,10,11\}$.  
\hpierre{To define $g$ and the Jacobian, we need to decide which path we remove. 
  Which one did you remove?   We should give the function $g$ for the first case, as an illustration.}%
Assuming that we remove the path $\cP_4$ and select $\cJ_1$, we obtain
$g(\bU) = (g_1(\bU),\dots,g_5(\bU))^t$ where 
$g_p(\bU) = L_{p}$ for $p=1,2,3$ and $g_p(\bU)=L_{p+1}$ for $p=4,5$,
and the Jacobian is computed by interpreting the $g_p(\bU)$ as functions of $U_1,U_2,U_3,U_6,U_8$ 
alone, with the other $U_j$'s fixed.  
The GLR-U density estimator in (\ref{eq:glr-u}) turns out to be
\hflorian{I re-checked with the formula you give in this paper (assuming the derivative)
by $\bu$ is in fact by $u_j$) and it gives the same. I also produced plots of the densities 
with each estimator and compared them. I am not entirely sure what you wanted the estimators
to look like, but I went for the shorter version and put \eqref{eq:glru2} in the same 
format as \eqref{eq:glru1}.}%
\begin{equation}
 \label{eq:glru1}
  G_1(\bU;x) = -\II[h(\bY)\leq x]\,\,\sum_{j=1}^{3} j^{-1}\Phi^{-1}(U_j) 
      = -\II[h(\bY)\leq x]\,\,\sum_{j=1}^{3}j^{-2}(Y_j-10j)
\end{equation}
for $\cJ_1$ and 
\begin{equation}
 \label{eq:glru2}
  G_1(\bU;x) = -\II[h(\bY)\leq x]\,\,\sum_{j=10}^{11} j^{-1}\Phi^{-1}(U_j) 
      = -\II[h(\bY)\leq x]\,\,\sum_{j=10}^{11}j^{-2}(Y_j-10j)
\end{equation}
for $\cJ_2$, where $\Phi$ denotes the standard normal cdf.

\begin{table}[!htbp]
\centering 
\caption{Estimated values of $-\log_2(\MISE)$ with $n=2^{20}$ points and estimated $\MISE$ 
   rate $\hat\beta$ for various methods, for the shortest path example.
	 The 21.3 entry (for example) means that for the KDE with MC and $n=2^{20}$ points,
	 we have $\MISE \approx 2^{-21.3}$. }
\label{tab:shortestPath}
\begin{tabular}{l | c c | c c}
 \hline 
  		  	& \multicolumn{2}{c|}{MC} 			&  \multicolumn{2}{c}{RQMC} \\
 Method 	&  $-\log_2(\MISE)$  & ${\beta}$  & $-\log_2(\MISE)$  & $\hat{\beta}$ \\
 \hline
 KDE 				      & 21.3 	& 0.8 & 25.7 	& 0.96 \\
 CDE (blue cut)		& 24.7 	& 1.0 & 45.6 	& 2.12 \\
 CDE (orange cut)	& 29.1 	& 1.0 & 46.5 	& 1.66 \\
 LRDE 				    & 20.2 	& 1.0 & 27.8 	& 1.38\\
 GLR-U in \eqref{eq:glru1}
					        & 15.4 	& 1.0 & 23.2		& 1.29	\\
 GLR-U in \eqref{eq:glru2}
					        & 21.5  & 1.0 & 29.6	& 1.35 \\
\hline
\end{tabular}
\end{table}

Table~\ref{tab:shortestPath} summarizes our numerical results for this example, for all the methods.
It reports $-\log_2(\MISE{})$ for $n=2^{20}$ as well as the convergence rate 
exponent $\beta$ for MC and its (noisy) estimate $\hat\beta$ for RQMC.
%
\hpierre{Start with the most important and striking observations that we want to highlight,
 and do not say too much.  We just want to direct the attention to the most important and interesting.}%
\hflorian{I know, that part was a mess. It didn't have any goal or structure or whatsoever. I 
 wanted to re-do it but ran out of time.}%

We find that the CDE combined with RQMC outperforms all other methods by a wide margin.
Compared with the KDE with MC (the traditional approach), it reduces the MISE for $n=2^{20}$
by a factor of about $2^{25} \approx 32$ millions. 
The orange cut $\cL_2$ does better than the blue cut $\cL_1$, especially for plain MC.
This could appear surprising, because $\cL_2$ has fewer arcs, but the explanation is that 
the two arcs of $\cL_2$ have a much larger variance, so it pays off to hide them. 
Generally speaking, we want to select a conditioning that hides (or integrate out) variables 
that capture as much variance as possible. 
(For the blue cut, the noise in the linear regression model and the estimate $\hat\beta$ 
appears quite significant.)

We also observe a significant difference of performance between the two choices of input
variables for GLR-U.  With $\cI_2$, the performance is better than for the KDE,
whereas for $\cI_1$ it is worse.
This shows that the choice of input variables may have a significant impact on the performance
in general.  
Note that $\cI_2$ contains input variables that have much more variance than $\cI_1$.
By comparing (\ref{eq:glru1}) and (\ref{eq:glru2}), we can see why the second estimator 
has less variance: the terms in the sum that multiplies the indicator have larger constants in 
the denominator, and therefore a smaller variance.
In some sense, the GLR-U estimator integrates out part of the variance contained in the 
selected input variables, so it makes sense to select a subset of input variables 
that captures more of the variance.

The LRDE has a larger $\MISE$ than the KDE with $n=2^{20}$ MC samples,
but it beats the KDE when using RQMC.
It also performs better than GLR-U for one choice of inputs and worse for the other choice.

With the same network, we now consider a slightly different problem.
We assume that the $Y_j$'s are random link capacities instead of random lengths,
and we want to estimate the density of the maximum flow that can be sent from the source to the sink. 
This maximum flow $h(\bY)$ is equal to the capacity of the minimal directed cut 
having the smallest capacity.
Here, we assume that $Y_j$ is normal with mean $\mu_j=10$ and standard deviation 
$\sigma_j=1$ for $j < 10$ and normal with mean $\mu_j=20$ and standard deviation 
$\sigma_j=4$ for $j =10$ and 11.
For the CDE, if we take $\cG$ as in the previous case, the distribution of $X$ conditional on $\cG$
typically has a probability mass at some point. 
For instance, if $\cL = \cL_1$, then after the conditioning, $Y_{10} + Y_{11}$ is known
and there is a positive probability that this is the value of the maximum flow.  
As a result, the conditional cdf is sometimes discontinuous and 
the CDE is no longer an unbiased density estimator.
This motivates the use of LRDE for this example.

Similarly as in the previous example, $h(\bY)$ is the minimum over several simple sums of $Y_j$'s,
so multiplying all $Y_j$'s by a positive constant multiplies the maximum flow $h(\bY)$ 
by the same constant. Therefore, the change of variables $\varphi_x(\bz)=x\bz$ can be used
again and provides the exact same LRDE as in \eqref{eq:lrde_paths}, but with the modified $h$, 
$\mu_j$, and $\sigma_j$.

\new{For GLR-U, the construction is similar as for the previous example, except that we 
select a subset of minimal cuts with independent capacities instead of a subset of paths.
There are hundreds of thousands of ways of selecting the subset of minimal cuts.
We tried a few of them and obtained the best results by selecting the set of cuts:
$\{ \{10,11\}, \{1,2,7\},\{1,2,9\},\{1,2,11\}, \{1,5,9\}, \{2,3,4\}, \{4,5,11\}, 
\{8,9,10\},\? \{2,3,\? 8,10\},\? \{6,7,8,10\} \}$ and then hiding $Y_{11}$.
This gives the estimator 
\begin{equation*}
 G(\bU;x) = -\II[h(\bY)\leq x] \left( (Y_1-10) + (Y_4-10) + (Y_{10} - 20)/16 \right).
\end{equation*}}

\new{Numerical results for the KDE, LRDE, and GLR-U for this example are given in Table~\ref{tab:maxFlow}.
In terms of \MISE, under MC, the LRDE performs better than GLR-U and slightly better than the KDE, but not much.
However, RQMC improves the \MISE{} for $n=2^{20}$ by a factor of about 30 for the LRDE, 
a bit more for GLR-U, and about 3 for the KDE.  
The combination of LRDE or GLR-U with RQMC also improves the convergence rate $\hat\beta$.}

\begin{table}[!htbp]
\centering 
\caption{Values of the $\log_2(\MISE)$ estimated with $n=2^{20}$ points and the 
  estimated $\MISE$ rate $\hat\beta$ for various methods for the maximum flow example.}
\label{tab:maxFlow}
\begin{tabular}{l | c | c c}
 \hline 
  			    & MC 						    &\multicolumn{2}{c}{RQMC} \\
 Method 	  &  $-\log_2(\MISE)$ & $-\log_2(\MISE)$  & $\hat{\beta}$ \\
 \hline
 KDE 				& 18.3 	& 19.7 	& 0.86 \\
 LRDE 		  & 18.7 	& 23.6 	& 1.23 \\
 GLR-U 		  & 17.7 	& 23.2 	& 1.26 \\
 \hline
\end{tabular}
\end{table}

\section{Conclusion}
\label{sec:conclusion}

We discussed and compared several recent developments regarding density estimation 
for simulation models, with Monte Carlo and quasi-Monte Carlo methods. 
Most of these methods provide unbiased density estimators and some of them are also 
RQMC-friendly, in which case their MISE can converge at a faster rate than the 
canonical rate of $\cO(1/n)$ as a function of the sample size $n$.
For the classical density estimators in statistics, in contrast, the MISE converges
at a slower rate than $\cO(1/n)$.
In our numerical example (and several other experiments not reported here),
the CDE combined with RQMC was by far the best performer.  
However, for some types of problems it may be difficult to apply, and then one can
rely on one of the alternatives.  
In future work, these density estimators should be adapted, implemented, 
and compared for a larger variety of Monte Carlo applications for which density 
estimates are useful.

\begin{acknowledgement}
This work has been supported by a NSERC Discovery Grant and an IVADO Grant to P. L'Ecuyer.
F. Puchhammer was also supported by Spanish and Basque governments fundings through BCAM (ERDF, ESF, SEV-2017-0718, PID2019-108111RB-I00, PID2019-104927GB-C22, BERC 2018e2021, EXP. 2019/00432, ELKARTEK KK-2020/00049), and the computing infrastructure of i2BASQUE academic network and IZO-SGI SGIker (UPV).
\end{acknowledgement}
\vskip-25pt

%

\end{document}